\documentclass[runningheads]{llncs}
\usepackage{graphicx}
\usepackage{booktabs}
\usepackage{amsmath}
\usepackage{array}
\usepackage{multirow}
\usepackage{hyperref}
\hypersetup{hidelinks}
\usepackage{url}
\usepackage{caption}

\title{Qubes OS Security in the Public Record}
\titlerunning{Qubes OS Security in the Public Record}
\author{Alfonso De Gregorio}
\authorrunning{Alfonso De Gregorio}
\institute{Pwnshow}

\begin{document}
\maketitle

\begin{abstract}
Qubes OS is a revealing case for security measurement because its architecture makes component boundaries security-relevant. We present a protocol-driven longitudinal analysis of 109 public Qubes Security Bulletins (QSBs, 2011--2025), the official Qubes-maintained Xen Security Advisory (XSA) tracker, and a secondary vulnerability-event sensitivity series. The study measures the public advisory record rather than latent vulnerability incidence or realized compromise. The methodology combines audited deterministic component attribution, change-point analysis, overdispersion checks, severity-proxy weighting, censoring sensitivity, documentary latency lower bounds, and baseline-aware evaluation of vulnerability discovery models (VDMs).

The results show persistent upstream dependence in that public record. On the official tracker, 113 of 464 XSAs affect Qubes; under primary labeling, 87 of 109 QSBs (79.8\%) are attributable to Xen, CPU/micro\allowbreak architectural, or other upstream components rather than Qubes-core logic, with similar results under weighted views. Change-point analyses identify 2015Q1 as the dominant break in the quarterly advisory series, while post-2018 annual disclosure rates are statistically flat. Poisson inferences are stable under dispersion diagnostics and negative-binomial sensitivity checks. The attribution codebook performs well in a stratified 30-QSB audit, and S-shaped VDMs fit descriptively but do not significantly outperform a rolling-mean baseline in short-horizon forecasts.

Overall, the Qubes public advisory record appears stable, but not quiet: disclosure activity plateaus at a higher level than in the earliest years, while the observed burden remains concentrated in upstream trust anchors.
\end{abstract}

\section{Introduction}

Qubes OS is built around security-by-compartmentalization. Rather than treating the desktop as one protection domain, it decomposes applications and major subsystems into separate virtual machines (``qubes'') running on the Xen hypervisor, with network and USB stacks intentionally pushed out of \texttt{dom0}. The official documentation is explicit that a limited set of interfaces remains security-critical, including Xen itself, device backends, and Qubes-maintained control paths such as \texttt{gui-daemon} and \texttt{qrexec-daemon}. The same documentation also emphasizes that the GUI virtualization subsystem adds only about 2{,}500 lines of C code to the privileged domain, illustrating the project's long-standing strategy of minimizing, rather than denying, trust in the residual TCB \cite{arch}.

Throughout the paper, we use \emph{Qubes-core} as the component label for Qubes-maintained control and policy logic such as \texttt{qrexec}, GUI virtualization, and closely related management paths. Phrases such as ``Qubes-specific control logic'' or ``Qubes-originated isolation failures'' are used only to emphasize mechanism or failure mode within that same broad class; they do not denote a separate fifth category.

This architecture makes Qubes OS analytically valuable. If compartmentalization is working as intended, one should expect relatively little Qubes-core advisory activity compared with advisories rooted in Xen, the processor execution substrate, or other upstream dependencies. At the same time, any measurement study in this space must be explicit about its object of analysis. This paper measures the \emph{public advisory record}: signed Qubes Security Bulletins (QSBs), the Qubes-maintained Xen Security Advisory (XSA) tracker, and a secondary sensitivity series derived from public crosswalks. It does \emph{not} directly measure latent vulnerability incidence, realized exploitation, or operational compromise rates. Disclosure policy, upstream reporting practice, advisory bundling, research attention, and tracking-process changes all mediate what becomes visible in that record.

Qubes publishes signed Qubes Security Bulletins (QSBs) \cite{qsb}, while Xen vulnerabilities are separately tracked as Xen Security Advisories (XSAs) \cite{xen-xsa}. The Qubes project also maintains a dedicated XSA tracker identifying which XSAs affect Qubes security \cite{xsa-tracker}. This yields an unusually transparent public record spanning more than fourteen years. Yet a rigorous measurement study still has to solve several known problems: advisories are not vulnerabilities, cumulative $R^2$ is optimistic, breakpoint selection can drift into narrative convenience, component attribution is rarely naturally single-label, and advisory counts are often modeled as Poisson without checking whether the data warrant that assumption.

Our goal is therefore twofold. Substantively, we ask what the Qubes public advisory record says about upstream dependence, regime changes, and the relative visibility of Qubes-core versus non-Qubes-core issues. Methodologically, we ask how far one can push an advisory-based longitudinal design without blurring the line between the public record and the underlying security state it only imperfectly reflects. In that sense, the paper is not an attack on prior Qubes commentary or on classical VDM work. Instead, it operationalizes the caution urged by the Massacci--Nguyen validation lineage on a case where architectural boundaries are unusually meaningful and the public record is unusually clean.

The paper contributes three things. First, it packages a reproducible QSB ledger, XSA-tracker snapshot, and associated analysis tables into a public artifact bundle. Second, it demonstrates a protocol for advisory-based measurement that combines auditable multi-scheme attribution, change-point analysis, overdispersion checks, censoring sensitivity, and baseline-aware forecast evaluation. Third, it reports a negative predictive result that is itself important: S-shaped VDMs remain descriptively useful on the Qubes series, but they do not earn a clear short-horizon forecasting advantage over strong rolling baselines.

The central substantive finding is narrower than a claim about ``Qubes security'' in the abstract. In the public advisory record, the burden is concentrated in upstream hypervisor, hardware, and adjacent upstream components rather than in Qubes-core logic. The quarterly record exhibits a dominant shift around 2015Q1, and the annual series after 2018 is best described as stable rather than accelerating. Those are claims about the published record and its architecture-aware decomposition; they are not direct claims about latent vulnerability incidence or about a fully observed operational security posture.

\section{Related Work and Architectural Context}

Classical VDM work adapted software reliability ideas to vulnerability discovery, often using non-homogeneous Poisson process (NHPP) or logistic-shaped cumulative curves. Alhazmi, Malaiya, et al. popularized the logistic framing for operating-system vulnerabilities \cite{alhazmi2005modeling,alhazmi2007}, while independent validation work by Nguyen and Massacci \cite{nguyen2012} and by Massacci and Nguyen \cite{massacci2014} showed that apparent goodness-of-fit can be misleading and that predictive utility must be evaluated directly. Ozment and Schechter's \emph{Milk or Wine} \cite{ozment2006} study of OpenBSD is especially relevant because it distinguishes architectural improvement from simple count decline. Beyond VDMs, the broader disclosure-and-patching literature has long warned that raw vulnerability counts are shaped by disclosure incentives, patch timing, and reporting practice as much as by latent software quality \cite{rescorla2005,arora2006}. More recent ecosystem studies of Android update delivery make the same point in a different advisory environment: the observability of security maintenance depends materially on release engineering and publication policy, not only on vulnerability incidence \cite{jones2020}.

Qubes OS adds a systems-security dimension absent from many earlier case studies. Its attack surface is intentionally redistributed across Xen, hardware, device domains, and a comparatively small amount of Qubes-core logic. That makes advisory counts semantically heterogeneous in a useful way: Xen and CPU bulletins speak to upstream dependence, while \texttt{qrexec}, GUI, and policy bulletins speak to the comparatively small amount of privileged Qubes-maintained logic. The official XSA tracker also applies an explicit relevance filter: a Xen issue is not marked as affecting Qubes when it is purely a host DoS, which is important for the interpretation of the public record.

Scheduler attacks in Xen-style environments show that shared-hypervisor resource management can create security-relevant failure modes even absent classical memory corruption \cite{zhou2011}. The transient-execution literature demonstrates that speculative and microarchitectural flaws are variant-rich, long-lived, and not fixable solely in software \cite{canella2019,trippel2018}. Detection and defense proposals based on hardware performance counters are promising but still incomplete \cite{ahmad2020,kosasih2024,precache2025}. At the same time, alternative isolation architectures such as Edera show that ``trusted anchor compression'' is not unique to Qubes but a general strategy for building strong isolation systems \cite{edera2025}. Qubes is therefore best understood as one instance of a broader pattern: isolation-heavy systems narrow the TCB they own directly, but become unusually dependent on the security economics of a few upstream anchors.

\section{Data and Protocol}

\subsection{Primary and secondary datasets}

The \emph{primary} dataset is a complete reconstruction of the public QSB index from QSB-001 (2011-05-12) through QSB-109 (2025-08-14), yielding 109 bulletins. This is the basis for all headline claims because it is directly reproducible from public sources. The official XSA tracker reports 464 XSAs from 2011-03-14 through 2026-01-27, of which 113 affect Qubes security. The tracker statistic is analytically important because it quantifies upstream dependence using the project's own relevance rules. The quarterly series therefore begins in 2011Q2, the first quarter containing a public QSB, and ends at the observed 2025Q3 partial endpoint. Because 2025 is right-censored relative to the preceding full years, every annual rate model is re-estimated with 2025 removed as an explicit sensitivity check.

The \emph{secondary} dataset is an annualized vulnerability-event sensitivity series of 147 issues through mid-2025, reconstructed from public QSB/XSA crosswalks and legacy working materials. Because event-level provenance for that series is incomplete in the public release bundle, it is used only for sensitivity analysis and not for primary causal or predictive claims.

We use three analytical units. The advisory unit is one QSB. The interval-count unit is quarterly or annual counts of QSBs. The secondary event unit is the annual aggregate vulnerability-event series. This layered design avoids pretending that bulletins and underlying vulnerabilities are the same object while preserving a fully reproducible main series.

\subsection{Attribution codebook and post hoc validation}

Each QSB was assigned one primary component label and one multi-label incidence set. The four categories are \textbf{Xen/hypervisor}, \textbf{CPU/microarchitecture}, \textbf{Qubes-core}, and \textbf{Upstream integration}. The codebook is deterministic and title-driven, using explicit tokens such as \emph{Xen}, \emph{XSA}, \emph{microcode}, \emph{speculative}, \emph{qrexec}, \emph{GUI}, \emph{Salt}, \emph{RPM}, and \emph{Linux netback}. This maximizes auditability, but deterministic coding can itself be biased, especially for mixed or editorially compressed bulletin titles. To assess reliability, we use a stratified 30-QSB post hoc validation subset. The subset includes all ten advisories whose titles triggered more than one label, plus twenty single-label advisories drawn by stratified random sampling across categories (6 Qubes-core, 5 CPU/microarchitecture, 5 Xen/hypervisor, and 4 upstream integration). The resulting audit set contains 8 Xen/hypervisor, 8 CPU/microarchitecture, 8 Qubes-core, and 6 upstream-integration bulletins, of which 10 are mixed-title cases. Manual validation was performed against the full bulletin text and linked identifiers without consulting the deterministic outputs during the initial pass; disagreements were then inspected to identify where the title-only rule set is most fragile.

To make attribution robustness explicit, we report four views of component burden:
\begin{enumerate}
\item \textbf{Primary}: one advisory contributes to one category.
\item \textbf{Incidence}: one advisory contributes one count to each category it touches.
\item \textbf{Weighted}: one advisory contributes $1/k$ to each of its $k$ categories.
\item \textbf{Identifier-weighted}: one advisory is weighted by the number of explicit XSA/CVE mentions in its title (minimum weight 1), then assigned by primary label.
\end{enumerate}
For simple binomial proportions we report Wilson confidence intervals; for weight\-ed and identifier-weighted shares we report percentile bootstrap intervals \cite{wilson1927}.
The fourth view is not a substitute for full severity enrichment; it is a transparent multiplicity proxy intended to test whether upstream dominance survives a coarse risk-weighting scheme.

\subsection{Trend, change-point, and count-model robustness}

Annual trends are first summarized with the classical Mann--Kendall rank-based trend test and Sen's median-slope estimator, which provide a distribution-light summary of monotonic trend in small annual series \cite{mann1945,kendall1975,sen1968}. Regime structure is then analyzed on quarterly counts using Bayesian single change-point inference, BIC-optimal piecewise Poisson segmentation, a first-split Poisson deviance search, and binary segmentation with a minimum segment length of four quarters. In the Bayesian one-break model, the break prior is uniform over admissible quarters, each segment count is Poisson-distributed, and the two segment rates have independent $\mathrm{Gamma}(1,1)$ priors; posterior mass is obtained by exact enumeration over candidate break locations.

To test the stable-regime claim directly, we fit a piecewise annual Poisson regression with an architecture-informed 2018 breakpoint:
\begin{equation}
Y_t \sim \mathrm{Poisson}(\lambda_t), \qquad \log \lambda_t = \beta_0 + \beta_1 t + \beta_2 (t-\tau)_+,
\end{equation}
where $\tau=2018$ and $(x)_+=\max(x,0)$. The pre-break log-rate slope is $\beta_1$, the slope change at 2018 is $\beta_2$, and the net post-break slope is $\beta_1+\beta_2$. The 2018 breakpoint is not data-mined: it brackets the Qubes 4.x architectural transition, including the simplification of VM classes and the shift away from legacy PV assumptions toward PVH/HVM-oriented defaults documented in the Qubes 4.0 release notes, core-stack announcement, and FAQ \cite{qubes40,core3,faq}. Because Poisson models can be fragile under overdispersion, we report Pearson dispersion and deviance/df, and fit negative-binomial robustness models. 

The robustness specification uses an NB2 mean-variance form, $\mathrm{Var}(Y_t)=\mu_t+\alpha\mu_t^2$, estimated by maximum likelihood; because $\alpha$ sits near the boundary in these data, we also report one-sided profile-likelihood upper bounds \cite{hilbe2011}. If variance materially exceeded the mean, the stable-regime interpretation would need re-evaluation; if not, Poisson remains adequate.

\subsection{Severity proxies and documentary latency lower bounds}

Bulletin-level CVSS enrichment is inconsistent, and many QSBs bundle multiple underlying issues. Rather than force a weak pseudo-CVSS panel, we use two transparent severity proxies. First, the identifier-weighted scheme above tests whether categories with more explicit XSA/CVE references dominate even when bulletins are weighted by issue multiplicity. This is intentionally coarse: it treats identifier multiplicity as an editorially observable burden proxy, not as a literal risk score. Second, a transient-execution tag identifies advisories involving speculative execution, sampling, microcode, branch prediction, or closely related microarchitectural mechanisms.

The mitigation-latency component remains constrained by public artifact availability. A full historical reconstruction from package-repository timestamps was not feasible here. Instead, we audit a recent subset of official QSB texts and the official testing-policy documentation. This yields a documentary lower bound and policy-aligned latency interpretation, not a full empirical latency distribution.

\subsection{VDMs, rolling forecasts, and significance tests}

We retain the standard cumulative VDM families \cite{anand2025,goel1979,musa1984,yamada1983,alhazmi2005modeling}:
\begin{align}
\hat N_{GO}(t) &= a\left(1-e^{-bt}\right),\\
\hat N_{MO}(t) &= a\ln(1+bt),\\
\hat N_Y(t) &= a\left[1-(1+bt)e^{-bt}\right],\\
\hat N_{AML}(t) &=  \frac{B}{1+C e^{-\kappa t}}.
\end{align}
Here \(B\) is the asymptotic vulnerability count, \(C\) is an integration constant controlling the horizontal displacement of the logistic curve, and \(\kappa\) is the composite growth coefficient. In the original Alhazmi--Malaiya formulation, the exponent is written as \(-ABt\); thus, \(\kappa\) corresponds to the product \(AB\) rather than to the standalone AML learning-rate parameter \(A\).

Rolling one-step-ahead annual forecasts are produced from cumulative non-linear least-squares (NLS) fits after a minimum five-year training window and compared against naive last-value, rolling three-year mean, drift, and Poisson-trend baselines. Forecast differences are tested using absolute-error loss, one-step horizon $h=1$, an effective comparison sample of ten annual forecast errors per series, and the Harvey--Leybourne--Newbold small-sample correction to the Diebold--Mariano test for equal predictive accuracy \cite{diebold1995,harvey1997}. To strengthen descriptive inference, we also fit annual-count Poisson increment likelihood models for Goel-Okumoto (GO), Musa-Okumoto (MO), Yamada, and Alhazmi--Malaiya Logistic (AML). These likelihood fits are used to report model-selection sensitivity and bootstrap parameter intervals for the two leading descriptive models: Yamada on the primary annual advisory series and AML on the secondary vulnerability-event series.

\section{Results}

\subsection{Exact disclosure dynamics and regime shifts}

\begin{figure}[t]
\centering
\includegraphics[width=.98\linewidth]{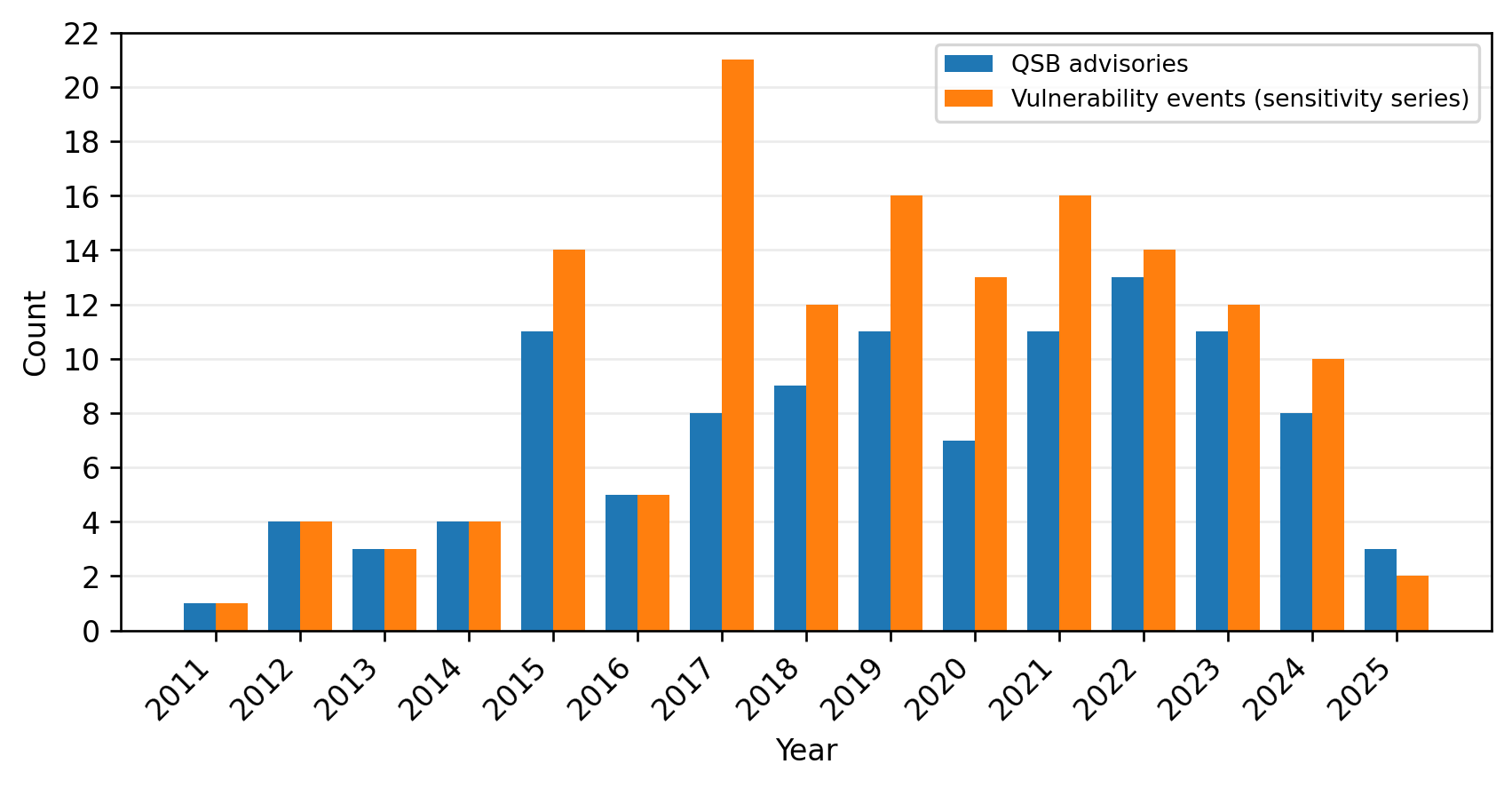}
\caption{Exact annual counts reconstructed from the canonical advisory tables. The primary series counts QSBs; the secondary series is the annualized sensitivity dataset of vulnerability events.}
\label{fig:annual}
\end{figure}

Figure~\ref{fig:annual} reports the annual series generated directly from the canonical CSV tables. The advisory record begins sparsely (1, 4, 3, and 4 QSBs in 2011--2014), shifts upward in 2015, and then remains active but bounded through the Qubes 4.x era. The secondary vulnerability-event series shows the same broad structure, with a sharper peak in 2017.

\begin{figure}[t]
\centering
\includegraphics[width=.98\linewidth]{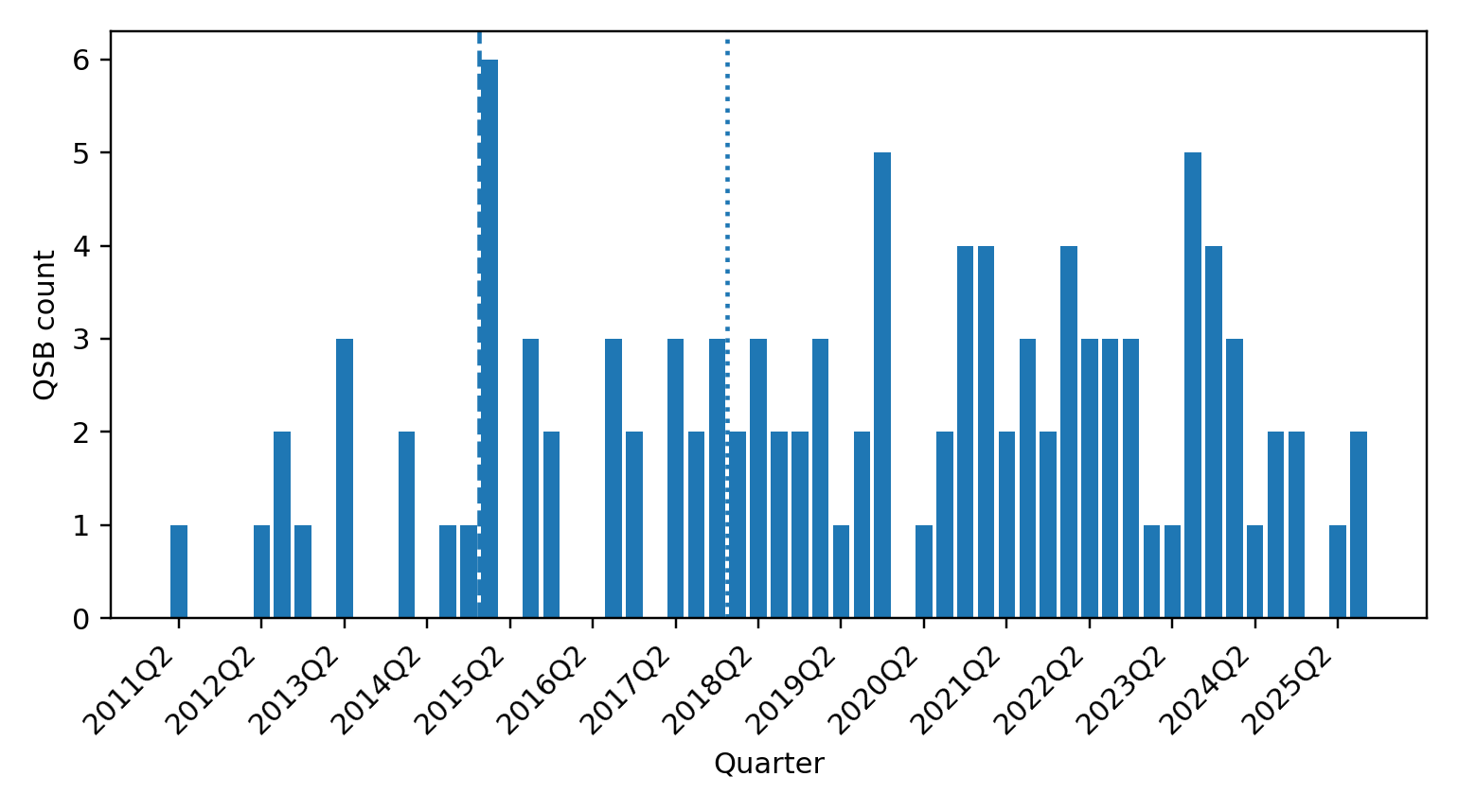}
\caption{Quarterly QSB counts with the dominant 2015Q1 break and the architecture-aware 2018 breakpoint. The 2015Q1 split is selected by Bayesian, BIC-based, deviance-search, and binary-segmentation analyses.}
\label{fig:quarterly}
\end{figure}

The nonparametric trend results are straightforward. The full annual QSB series shows a significant upward trend from the tiny early baseline, but the post-2018 period does not. Formal change-point analysis points consistently to 2015Q1 as the dominant break, and Fig.~\ref{fig:quarterly} shows the corresponding quarterly series. The Bayesian single change-point and BIC-optimal one-break Poisson segmentation both identify 2015Q1 as the dominant break. The BIC-optimal split is 2015Q1 (BIC 195.18), ahead of the nearest alternatives 2014Q4 (196.00) and 2014Q3 (196.78). Under the Bayesian one-break model, 2015Q1 is likewise the posterior mode with 25.7\% posterior mass; most of the remaining mass is concentrated in adjacent late-2014 quarters, and the combined posterior mass on the 2014Q3--2015Q1 window is 56.3\%. A quarterly Poisson deviance split search reaches the same answer, and binary segmentation with a four-quarter minimum segment length also selects 2015Q1 as the first split. Because the quarterly series is still short in absolute terms, these procedures should be read as convergent evidence for a segmentation of the public record rather than as proof of a unique underlying causal break. Secondary splits appear only much later and do not change the interpretation of the post-2018 period as a stable regime rather than a new growth regime.

The count-model robustness checks show that this inference is not a Poisson artifact. For the annual piecewise model including the partial 2025 endpoint, Pearson dispersion is 1.16 and deviance/df is 1.15; for the quarterly regime model around 2015Q1, the corresponding values are 0.97 and 1.18. In both cases, the NB2 overdispersion parameter collapses essentially to zero ($\hat\alpha \approx 8\times 10^{-8}$ annually and $\hat\alpha \approx 8\times 10^{-5}$ quarterly). One-sided 95\% profile-likelihood upper bounds are still small ($\alpha<0.15$ annually and $\alpha<0.21$ quarterly), corroborating the practical equivalence to Poisson for this dataset. The quarterly post-2015 rate ratio is 2.82 (95\% CI 1.55--5.14, $p=0.0007$), confirming that the early sparse regime and the later sustained regime are statistically distinct. The annual piecewise model supports a stable post-2018 public-record regime: the pre-2018 slope is positive, the slope-change term is negative, and the net post-2018 slope remains statistically indistinguishable from zero. Excluding the right-censored 2025 endpoint makes that conclusion stronger rather than weaker: the net post-2018 log-rate slope changes from $-0.058$ ($p=0.208$) in the full annual model to $-0.001$ ($p=0.989$) when only full years 2011--2024 are used, with Pearson dispersion falling to 0.88. The secondary annual vulnerability-event series likewise shows no significant post-2018 monotonic trend ($z=-0.92$, $p=0.356$ for 2018--2024), suggesting that advisory bundling does not drive the observed stabilization of the public record.

\subsection{Attribution robustness and validation}

\begin{figure}[t]
\centering
\includegraphics[width=.98\linewidth]{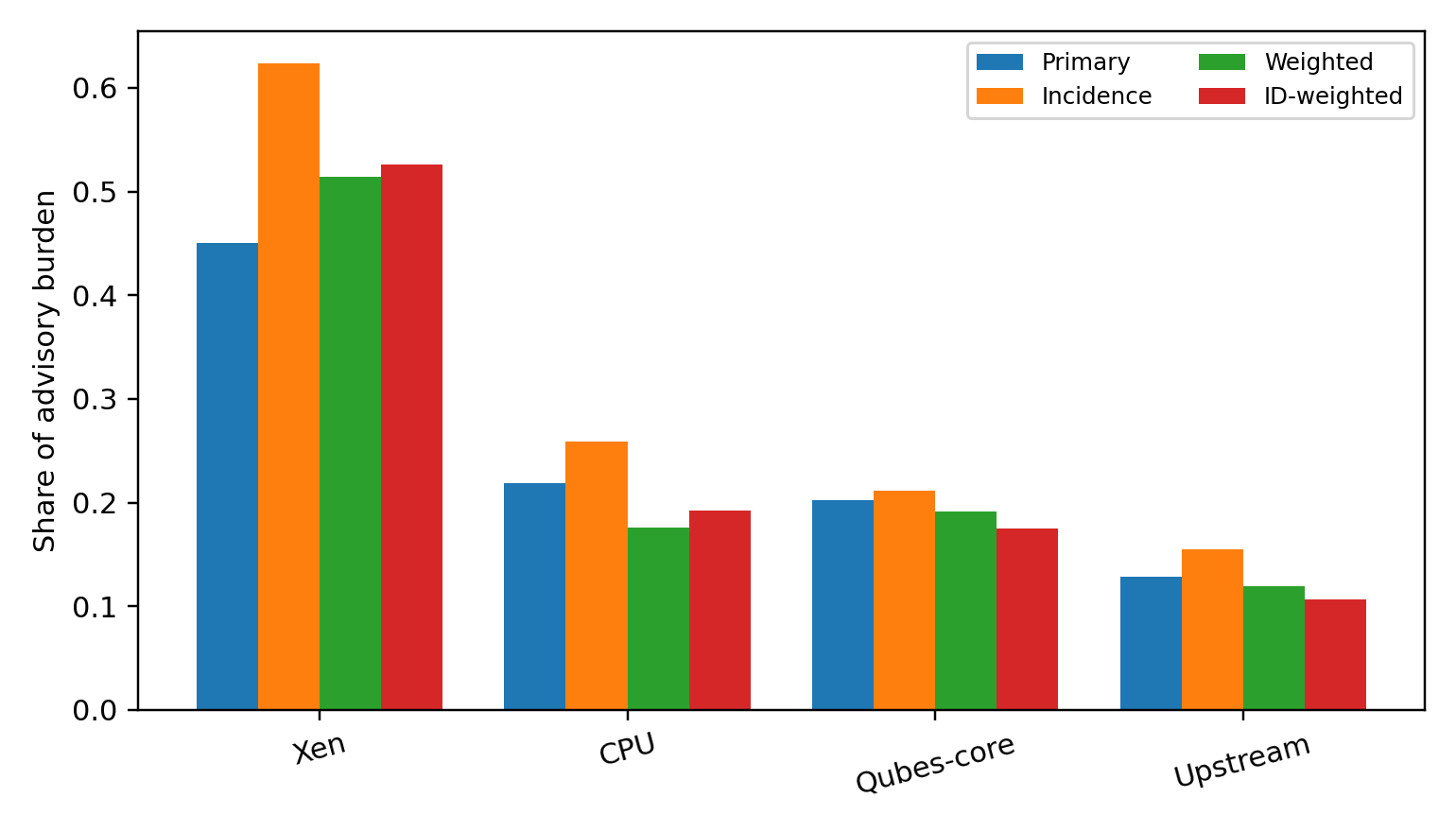}
\caption{Component shares under four attribution views. The conclusion that upstream components dominate is stable across schemes.}
\label{fig:shares}
\end{figure}

Table~\ref{tab:attrib} summarizes the four attribution views, and Fig.~\ref{fig:shares} shows the same result visually. For the incidence view, because categories are non-exclusive, the table reports upstream share as a share of total incidence mass rather than as a mutually exclusive advisory partition. Under the original primary scheme, 87 of 109 bulletins (79.8\%, 95\% CI 71.3--86.3) are attributable to Xen/hypervisor, CPU/micro\allowbreak architecture, or upstream integration rather than Qubes-core logic. This conclusion survives the more conservative weighted view (80.9\%, bootstrap 95\% CI 73.5--87.8) and becomes slightly stronger under identifier weighting (82.5\%, CI 75.2--89.1). Qubes-core remains a minority under every scheme: 20.2\% in primary counts, 19.1\% in weighted counts, and 17.5\% in identifier-weighted counts. Adjacent to the tracker-based statistic, this matters because the QSB-only attribution results point in the same direction even if one ignores the tracker's own relevance policy.

The post hoc validation subset suggests that the deterministic codebook is usable but not infallible. Primary-label accuracy is 96.7\% with macro-F1 0.969. For multi-label attribution, micro-F1 is 0.988 and the samplewise Jaccard index is 0.983. The only primary-label disagreement in the 30-QSB subset is a mixed GUI/Xen bulletin (QSB-034), which is exactly the sort of hybrid advisory for which deterministic single-primary assignment is most fragile. This is the right kind of error profile: the codebook appears strong on the many unambiguous titles and weaker only where the underlying bulletin itself is mixed.

\begin{table}[t]
\centering
\caption{Attribution robustness. ``Upstream share'' combines Xen/hypervisor, CPU/microarchitecture, and upstream integration. For the incidence view, the reported upstream share is the share of total incidence mass attributable to upstream categories. Weighted and identifier-weighted intervals are bootstrap intervals.}
\label{tab:attrib}
\small
\begin{tabular}{p{0.30\linewidth}rrp{0.28\linewidth}}
\toprule
View & Qubes-core & Upstream share & Notes \\
\midrule
Primary attribution & 20.2\% & 79.8\% & 87/109 upstream-coded advisories \\
Incidence attribution & 21.1\% & 83.1\% & incidence mass: Xen 62.4\%, CPU 25.9\%, Upstream 15.5\% \\
Weighted multi-label attribution & 19.1\% & 80.9\% & bootstrap 95\% CI 73.5--87.8 \\
Identifier-weighted attribution & 17.5\% & 82.5\% & bootstrap 95\% CI 75.2--89.1 \\
\bottomrule
\end{tabular}
\end{table}

\begin{table}[t]
\centering
\caption{Manual-validation audit summary for the deterministic attribution codebook.}
\label{tab:audit}
\small
\begin{tabular}{p{0.42\linewidth}p{0.46\linewidth}}
\toprule
Validation target & Result \\
\midrule
Primary labels & Accuracy 96.7\%, macro-F1 0.969 \\
Multi-label attribution & Micro-F1 0.988, samplewise Jaccard 0.983 \\
Observed primary disagreement & One mixed GUI/Xen bulletin (QSB-034) \\
\bottomrule
\end{tabular}
\end{table}

\subsection{Severity proxies, transient execution, and upstream pressure}

The severity-proxy analysis points in the same direction. Weighting advisories by explicit XSA/CVE mention count increases the Xen share to 52.7\% and leaves the combined upstream share at 82.5\%. This does not turn identifier count into a risk score; it shows that even when bulletin granularity and explicit issue multiplicity are given more weight, the same architectural conclusion survives. The transient-execution tag is also revealing. Twenty-three of the 109 QSBs (21.1\%) are transient-execution or microcode-related, and 23 of the 24 CPU/microarchitecture bulletins fall into this class. All transient-execution-tagged advisories occur from 2018 onward, and they account for 23 of 73 post-2018 bulletins (31.5\%).

This matters because it ties the post-2018 composition of the public Qubes advisory record to external research waves rather than to a simple internal quality narrative. The CPU-heavy years are not evidence that Qubes suddenly became less disciplined; they are evidence that hardware-mediated isolation erosion remained a live research and disclosure frontier long after the earliest Spectre/Meltdown wave \cite{canella2019,trippel2018}. In other words, the plateaued disclosure regime is not homogeneous. Its composition shifts toward microarchitectural risk precisely where the broader literature says it should.

\subsection{Documentary latency lower bounds}

A full public timestamp reconstruction of advisory-to-testing and testing-to-stable latency was not feasible from uniformly available public artifacts, but the documentary record still supports a bounded operational interpretation. We audited five representative official bulletins spanning 2019--2025 (QSB-055, QSB-083, QSB-089, QSB-102, and QSB-109) \cite{qsb055,qsb083,qsb089,qsb102,qsb109}. In all five cases, the bulletin text states that the relevant packages are already available in the \texttt{security-\allowbreak testing} repository at bulletin publication. In all five cases, the bulletin text also states that migration to the stable repository is expected over roughly two weeks after community testing. The official Qubes testing-policy documentation independently states that security updates are first uploaded to \texttt{security-testing} and, in general, remain there for about two weeks before migrating to \texttt{current}, with immediate stable release reserved for exceptional cases \cite{testing}.

This does not amount to a historical latency distribution, and we do not present it as such. Public repository history is heterogeneous across release eras, so mixing spotty timestamps into a pseudo-panel would be less reliable than retaining a clear documentary lower bound. In the audited subset, advisory-to-testing latency is explicitly zero days in the bulletin text, and the documented intended testing-to-stable window is approximately two weeks. That is enough to support a narrower operational claim than before: recent QSB practice is consistent with immediate publication into the security-testing channel and a roughly two-week policy-targeted stable migration cadence.

\subsection{VDMs: descriptive fit, likelihood sensitivity, and forecast inference}

The VDM analysis separates descriptive shape from predictive usefulness. On Poisson increment likelihood for annual counts, Yamada is the best descriptive model for the primary QSB series (AIC 72.06), clearly ahead of AML (77.08) and far ahead of GO and MO (89.6 each). For the secondary vulnerability-event series, AML is best (AIC 86.77), followed by Yamada (93.55), with GO and MO again much worse. Thus, even under likelihood-based estimation rather than cumulative $R^2$, the same high-level descriptive conclusion survives: the annual count series is better captured by S-shaped families than by simple exponential/logarithmic families.

Bootstrap intervals around the leading descriptive models are wide but informative. For the primary series, Yamada's asymptote parameter is $a=240.8$ with a 95\% bootstrap interval of approximately [159.7, 729.0], and the shape parameter is $b=0.102$ [0.046, 0.151]. For the secondary series, the AML fit yields \(\kappa=0.463\,[0.385,0.552]\), \(B=155.8\,[128.2,181.3]\), and \(C=62.1\,[33.8,142.2]\). We report \(\kappa\) here as the fitted composite exponent coefficient; in the original Alhazmi--Malaiya notation, this quantity corresponds to \(AB\). These intervals are too wide to justify a precise exhaustion claim, but they are still consistent with late-stage, slowing growth rather than open-ended acceleration.

\begin{figure}[t]
\centering
\includegraphics[width=.98\linewidth]{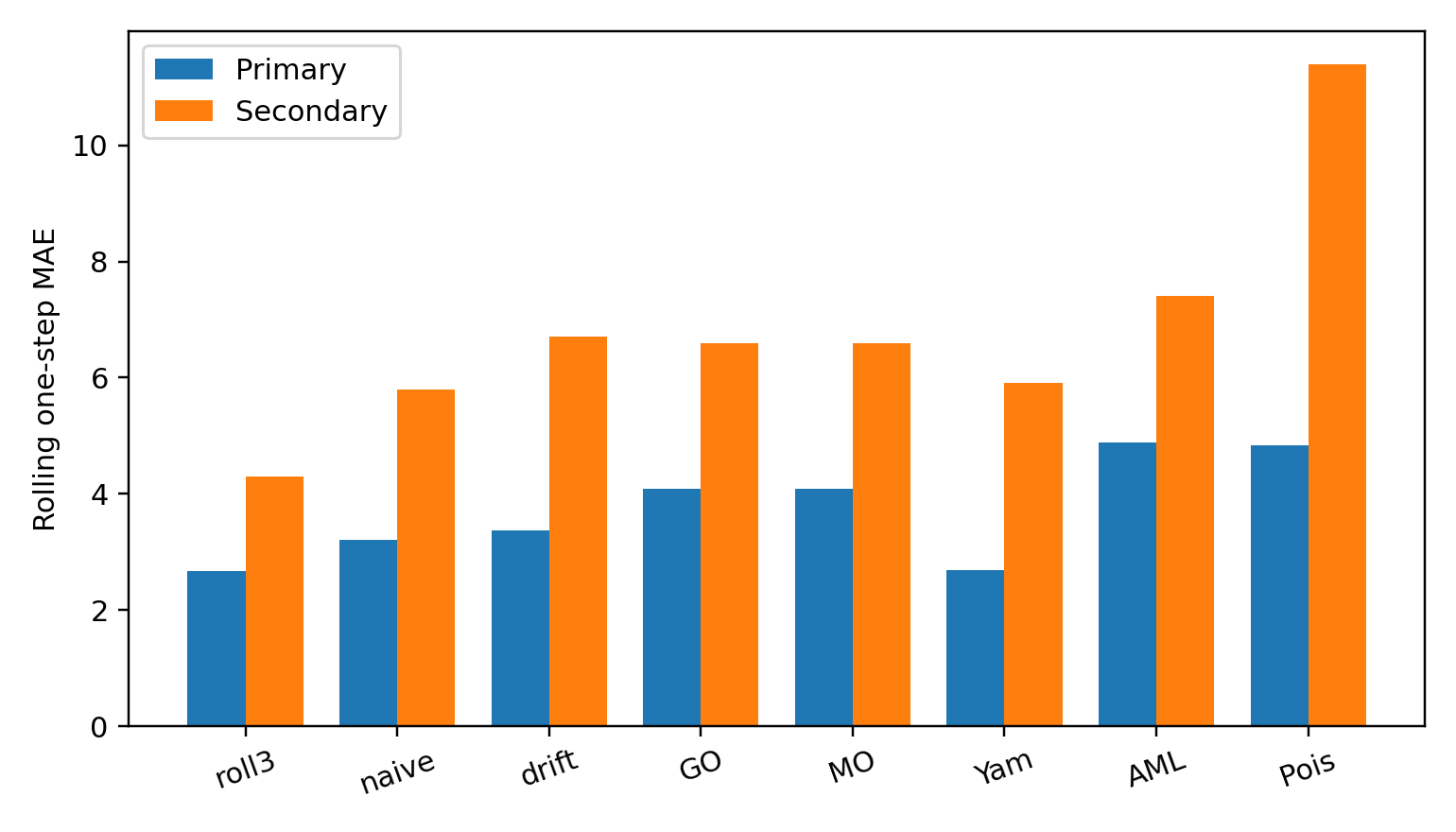}
\caption{Rolling one-step-ahead annual MAE. Simple rolling baselines remain hard to beat on both the primary and secondary series.}
\label{fig:mae}
\end{figure}

Forecasting remains the tougher test. Figure~\ref{fig:mae} and Table~\ref{tab:forecast} report rolling one-step-ahead annual forecasts after a minimum five-year training window. On the primary QSB series, the rolling three-year mean is best by MAE (2.67) and RMSE (3.32), with Yamada close in MAE (2.68) but not better. On the secondary series, the rolling three-year mean is again best (MAE 4.30), with Yamada second (5.91) and the remaining VDMs generally worse. Crucially, the Diebold--Mariano tests do not show a statistically significant advantage for the best VDM over the strongest baseline: on the primary series, rolling-three-year mean vs.~Yamada yields $p=0.978$; on the secondary series, the corresponding test yields $p=0.090$. Given the ten-error effective sample and the annual horizon, these tests are necessarily low-power; their role is therefore to rule out strong forecasting claims, not to prove exact equivalence. In other words, the data do not support the claim that VDMs provide a clear short-horizon forecasting edge over simple persistence-based baselines.

\begin{table}[t]
\centering
\caption{Rolling one-step-ahead annual forecast metrics and selected Diebold--Mariano (DM) tests. Lower MAE/RMSE is better.}
\label{tab:forecast}
\scriptsize
\begin{tabular}{p{0.20\linewidth}rrrr}
\toprule
Method & Primary MAE & Primary RMSE & Secondary MAE & Secondary RMSE \\
\midrule
Rolling 3-year mean & 2.67 & 3.32 & 4.30 & 5.78 \\
Naive last year & 3.20 & 3.52 & 5.80 & 7.27 \\
Drift & 3.38 & 4.15 & 6.70 & 8.20 \\
GO & 4.08 & 4.76 & 6.59 & 7.83 \\
MO & 4.08 & 4.76 & 6.59 & 7.83 \\
Yamada & 2.68 & 3.36 & 5.91 & 7.26 \\
AML & 4.88 & 5.89 & 7.40 & 9.06 \\
Poisson trend & 4.84 & 6.06 & 11.39 & 11.87 \\
\midrule
\multicolumn{5}{l}{\emph{Selected DM tests (absolute-error loss)}} \\
Rolling 3-year mean vs. Yamada & \multicolumn{2}{r}{Primary $p=0.978$} & \multicolumn{2}{r}{Secondary $p=0.090$} \\
Rolling 3-year mean vs. naive last & \multicolumn{2}{r}{Primary $p=0.495$} & \multicolumn{2}{r}{Secondary $p=0.215$} \\
Yamada vs. AML & \multicolumn{2}{r}{Primary $p=0.122$} & \multicolumn{2}{r}{Secondary $p=0.498$} \\
\bottomrule
\end{tabular}
\end{table}

The practical conclusion is therefore narrower than a classic VDM narrative. S-shaped families remain useful for descriptive shape and for stating that the Qubes advisory record is in a later stable phase rather than an explosive one. They do not earn the right to displace simple baselines for short-horizon operational forecasting.

\section{Discussion}

Three conclusions emerge.

First, the main architectural claim is robust across the analyses presented here. Upstream dependence is stable across primary, weighted, incidence, and identifier-weighted attribution, and the deterministic codebook performs well in a manual audit. Qubes-core advisories do occur, and some are serious, but they remain a minority of the public record. This is not proof that Qubes-core logic is intrinsically safe. It is evidence that the published QSB record is concentrated where the architecture says risk should be most visible: in Xen, in upstream integrations, and especially in the hardware execution substrate.

Second, the stable-regime interpretation survives stronger statistical scrutiny. The dominant break is not a narrative convenience but a repeatedly recovered 2015Q1 shift. The post-2018 period is not ``quiet,'' but it is stable enough that the net slope is effectively flat under the piecewise count model. Overdispersion checks do not rescue an alternative story. What changed after 2018 is less the level of visible activity than the composition of that activity: a larger share is explained by transient-execution and microcode waves, which are exogenous to Qubes-core implementation discipline. Because the annual and quarterly samples remain small, this should still be read as a cautious statement about the published record rather than as a strong causal diagnosis of underlying security dynamics.

Third, the VDM evidence is strongest when interpreted conservatively. A common temptation in advisory-based OS studies is to emphasize that an S-shaped curve fits well and then extrapolate toward exhaustion. The stronger result here is narrower but more credible: S-shaped curves remain the best descriptive family, but they do not clearly outperform strong naive baselines in short-horizon forecasts. That is exactly the kind of negative predictive result that a serious measurement paper should be willing to report.

These findings generalize beyond Qubes. Systems that pursue strong isolation by compressing trust into a small number of anchors often look similar: they can reduce the amount of security-critical logic they own directly, but they also become unusually exposed to the engineering quality and disclosure dynamics of their hypervisor, processor, and adjacent low-level ecosystem \cite{zhou2011,canella2019,edera2025}. Qubes is a particularly clean case because its public advisory corpus and architectural documentation make that dependency unusually observable. The contribution here is therefore not merely that ``upstream risk matters, which practitioners already expect, but that a public-record protocol can show how that dependence dominates the observable advisory stream, where regime changes appear, and which popular VDM claims fail to survive baseline-aware forecast checks.

\section{Limitations}

This study remains constrained by public advisory data. The primary QSB series is fully reproducible, but the secondary vulnerability-event series is public only as annual aggregates in the released artifact bundle. More fundamentally, the measured object is the published record itself. Disclosure policy, upstream reporting practice, advisory bundling, research attention, and tracking-process changes all influence what enters that record and when. The deterministic attribution codebook remains title-driven; the stratified full-text audit shows high agreement, but nuanced mixed advisories can still be editorially compressed in titles. The manual validation subset is a post hoc audit, not full independent dual-coder annotation of the entire corpus. The severity proxies are transparent and useful, but they are not a substitute for complete bulletin-level CVSS enrichment. The documentary latency analysis still falls short of a historical repository-timestamp reconstruction. The XSA tracker itself encodes a relevance policy, including the exclusion of pure host DoS cases; any historical drift in that policy could affect tracker-based upstream percentages, although the main upstream-dependence result does not rely on the tracker alone and survives QSB-only attribution. Finally, the annual forecast evaluation remains sample-limited; Diebold--Mariano tests on such short series have limited power, so ``no significant advantage'' should be read as a cautious result rather than a proof of equivalence. Right-censoring of the 2025 endpoint remains a real concern for any public advisory series, but the endpoint-exclusion sensitivity shows that this does not overturn the stable-regime inference here.

These limitations are conservative. They make it harder to overclaim precise exhaustion, exact patch-timing behavior, or latent vulnerability incidence, but they do not undercut the main claims about how the public advisory record is structured and how that structure reflects Qubes' upstream dependence.

\section{Data Availability}

A public replication bundle accompanies this preprint and is prepared for archival deposition. It contains the complete primary 109-QSB ledger, annual and quarterly count tables, the XSA-tracker summary snapshot used in the analysis, derived analysis tables, the deterministic attribution codebook, and the stratified manual-validation subset. Consistent with the release scope described in Sect.~3, the secondary vulnerability-event series is released at the annual aggregate level used for sensitivity analysis in this paper \cite{DeGregorio2026data}.

\appendix
\section{Deterministic Attribution Codebook (Summary)}

The public artifact bundle contains the full codebook and per-QSB assignments. This appendix records the summary logic used in the manuscript. The deterministic title-driven rules assign one primary component label and, where appropriate, multiple incidence labels. Representative triggers are as follows.

\begin{table}[t]
\centering
\caption{Summary of deterministic attribution rules. The public artifact contains the complete token list and precedence rules.}
\label{tab:codebook}
\small
\begin{tabular}{p{0.22\linewidth}p{0.70\linewidth}}
\toprule
Category & Representative title triggers \\
\midrule
Xen/hypervisor & Xen, XSA, libxl, grant table, PV, PVH, HVM, scheduler, hypervisor, shadow paging \\
CPU/micro\allowbreak -architecture & microcode, speculative, transient, branch prediction, Retbleed, Spectre, BHB, SRSO, sampling \\
Qubes-core & qrexec, GUI, clipboard, policy, parser, service request, Qubes daemon, core-admin \\
Upstream integration & Salt, RPM, Linux netback, kernel driver, template packaging, domU integration issue \\
\bottomrule
\end{tabular}
\end{table}

Mixed titles receive all triggered incidence labels. Primary labels follow a fixed precedence order intended only for summary statistics: CPU/microarchitecture first, then Xen/hypervisor, then Qubes-core, then upstream integration. The public artifact bundle records both the raw trigger matches and the final assigned labels so that readers can reproduce or revise the mapping.

\section{Conclusion}

This paper presents a protocol-driven longitudinal analysis of Qubes OS public advisories using a stricter design than is typical for advisory-based studies. The primary dataset --- 109 QSBs over 2011--2025 --- is fully reproducible. The analysis combines overdispersion robustness, formal change-point sensitivity, audited multi-scheme attribution, severity-proxy weighting, documentary latency lower bounds, and baseline-aware forecast inference.

The resulting picture is clear. The Qubes public advisory record enters a higher but stable regime after the dominant 2015Q1 break and remains statistically flat after 2018 in annual counts. Within that record, the dominant burden remains upstream rather than Qubes-core. S-shaped VDMs remain useful for describing the long-run curve, especially Yamada on the primary advisory series and AML on the secondary vulnerability series, but they do not earn a clear short-horizon forecasting advantage over simple rolling baselines.

For practitioners, the implication is more concrete than simply ``watch upstream.'' The evidence here suggests that assurance effort should follow Xen and CPU anchors operationally: monitor the Qubes XSA tracker and Xen advisory stream as first-class inputs, track microcode- and transient-execution-driven update waves, and treat the documented security-testing-to-stable cadence as a real exposure-management window when deciding how aggressively to stage or accelerate updates. For the measurement community, the main lesson is methodological: advisory-based studies of compartmentalized systems should test for overdispersion, audit their attribution rules, and evaluate forecasting value against naive baselines before turning descriptive curves into causal claims about security maturity.

\end{document}